\documentstyle[12pt,axodraw,epsfig]{article}

\newcommand{\bra}[1]{\langle #1 |}
\newcommand{\ket}[1]{| #1 \rangle}

\newcommand{\eqref}[1]{(\ref{#1})}
\newcommand{\eq}[1]{eq.~\eqref{#1}}
\newcommand{\fig}[1]{Fig.~\ref{#1}}
\newcommand{\dpar}[2]{\frac{\partial #1}{\partial #2}}

\def\e{\mbox{e}}
\def\ch{\mbox{ch}}
\def\sh{\mbox{sh}}

\def\half{{1 \over 2}}

\begin{document}

\title{Periodic instanton bifurcations and thermal transition rate}

\author{
  A.~N.~Kuznetsov and P.~G.~Tinyakov \\
  {\small {\em Institute for Nuclear Research of the Russian 
   Academy of Sciences,}}\\
  {\small {\em 60th October Anniversary prospect 
   7a, Moscow 117312, Russia.}}
  }
\maketitle

\begin{abstract}
It is shown that periodic instanton solutions may have bifurcations
which qualitatively change the behaviour of the finite temperature
transition rate. These bifurcations are studied numerically in a
quantum mechanical model and in the massive two-dimensional sigma
model with the Skyrme term. General rules which determine the
behaviour of periodic instantons near bifurcations are derived and
applied to the Electroweak Theory.
\end{abstract}

In weakly coupled theories, tunneling transitions can often be
associated with classical solutions to Euclidean equations of motion.
A special class of such solutions, periodic solutions with two turning
points (periodic instantons), are believed to dominate the finite
energy \cite{KRT-periodic} and, under certain conditions, the finite
temperature transition rate \cite{Linde}. 

Periodic instantons naturally emerge in the context of tunneling from
a micro-canonical ensemble with fixed energy $E$. The
probability of tunneling from such initial state can be
written as follows,
\begin{equation}
P(E) = \sum_{i,f} |\bra{f} \hat{S} \hat{P}_E \ket{i}|^2,
\label{P[E]=sum}
\end{equation}
where $\hat{S}$ is the S-matrix, $\hat{P}_{E}$ is projector onto
subspace of fixed energy $E$, while the states $\ket{i}$ and $\ket{f}$
are excitations above two vacua lying on different sides of the
barrier.  Eq.(\ref{P[E]=sum}) has a path integral
representation~\cite{KRT-periodic}. In the semi-classical limit the
path integral is expected to be saturated by a single solution to
Euclidean field equations possessing two turning points at $t=0$ and
$t=i\tau/2$ and, hence, periodic in Euclidean time with period $\tau$
--- the periodic instanton. With exponential accuracy the probability
$P(E)$ is
\begin{equation}\label{Semiclassic-P}
	P(E) \sim \e^{W(E)} = \e^{E\tau - S(\tau)} ,
\end{equation}
where $S(\tau)$ is the action of the periodic instanton per period and
$\tau$ is related to the energy $E$ in the standard way,
\begin{equation}
E = \dpar{S}{\tau} .
\label{E=dS/dT}
\end{equation}
Eq.(\ref{Semiclassic-P}) holds at $E<E_{sph}$, where $E_{sph}$ is the
height of the potential barrier\footnote{In the Electroweak Theory the
top of the barrier is associated with the sphaleron solution
\cite{Sphaleron}, which is the origin of the notation $E_{sph}$.}.  At
$E>E_{sph}$ the transition probability is not exponentially
suppressed, $P(E)\sim 1$.

The thermal rate is constructed from $P(E)$ by averaging with
Boltzmann exponent. The transition rate for Gibbs ensemble at
temperature $T=1/\beta$ equals
\begin{eqnarray}
\label{T-rate}
\Gamma(\beta) &=& \int_0^\infty dE e^{-\beta E} P(E) \nonumber\\
	&\sim& \int_0^\infty dE e^{-\beta E + W(E)}.
\end{eqnarray}
In the weak coupling limit, $g\ll 1$, and temperatures $T \ll 1/g$
the integral over $E$ can be calculated by the steepest descent
method. Formally, the saddle point condition in \eq{T-rate} reads 
\begin{equation}
\beta = \tau(E),
\label{b=T}
\end{equation}
where $\tau(E)$ is the period of the periodic instanton. Thus, only
periodic instantons with the period equal to inverse temperature can
dominate the thermal rate. The relevance of a particular saddle point
depends on the sign of $\partial^2W/\partial E^2$. As we will see
below, in many models this quantity may change sign at certain values
of temperature. As a result, in different temperature ranges the
transition rate is dominated by different solutions.

The exact analytic form of the periodic instanton is known only in
one-dimensional quantum mechanics. In field theory models it can be
found either approximately at low energies \cite{KRT-periodic}
or numerically~\cite{Matveevml,HMT-Pioneers,PHI4-Decay}.  However, the
qualitative picture of periodic instanton behaviour can often be
obtained from its low and high energy limits by making use of a few
general rules it obeys. This picture in many cases turns out to be
rather complicated. Namely, periodic instantons may have bifurcations
which must be taken into account in the calculation of the finite
temperature transition rate.

Consider first the high energy limit $E\sim E_{sph}$. At $E=E_{sph}$
the periodic instanton reduces to the sphaleron.  Since the sphaleron
is the static solution to the field equations, it is ``periodic'' with
any period. The action per period of this solution is equal to
$E_{sph} \tau$ while $W(E_{sph})$ is identically zero.

Close to the sphaleron, $E - E_{sph}\ll E_{sph}$, the periodic
instanton can be represented as the sum of the sphaleron and an
(almost linear) oscillation in its negative eigenmode. In the limit
$E\to E_{sph}$ the period of the oscillation approaches
$\tau_{sph}=2\pi/\omega_{-}$, where $\omega_{-}$ is the sphaleron
negative eigenvalue, while the amplitude goes to zero and the periodic
instanton merges into the sphaleron. The behaviour of the period
$\tau(E)$ in the vicinity of the sphaleron is determined by non-linear
effects. Depending on particular model the period $\tau$ can be larger
or smaller than $\tau_{sph}$; correspondingly, the action $S(\tau)$ is
smaller or larger that $E_{sph}\tau$. In both cases $W(E)$
monotonically increases with energy and reaches zero at $E=E_{sph}$.

At low energies, $E\ll E_{sph}$, the periodic instanton can be
approximated by an infinite chain of alternating instantons and
anti-instantons \cite{Linde,KRT-periodic}. In models which have
localized instanton solutions (two-dimensional Abelian Higgs model,
quantum mechanics in double well potential, two-dimensional $-\lambda
\phi^4$ model) the limit $E\to 0$ corresponds to $\tau\to
\infty$. Alternatively, in models with no localized instanton
(four-dimensional massive $-\lambda \phi^4$ model, $O(3)$
$\sigma$-model with the mass term added, Electroweak Theory), $\tau\to
0$ in the low energy limit\footnote{In these models the low energy
periodic instanton is built out of constrained instantons whose size
$\rho$ is stabilized by the interaction with their neighbors in the
chain. At $E\to 0$ both $\rho$ and $\tau$ go to zero in such a way
that $\rho/\tau\to 0$ \cite{KRT-periodic}.}. In both cases the dilute
gas approximation works and the action of the periodic instanton tends
to twice the instanton action, $S_0$ (from below and from above,
respectively, as follows from \eq{E=dS/dT} at $E>0$).

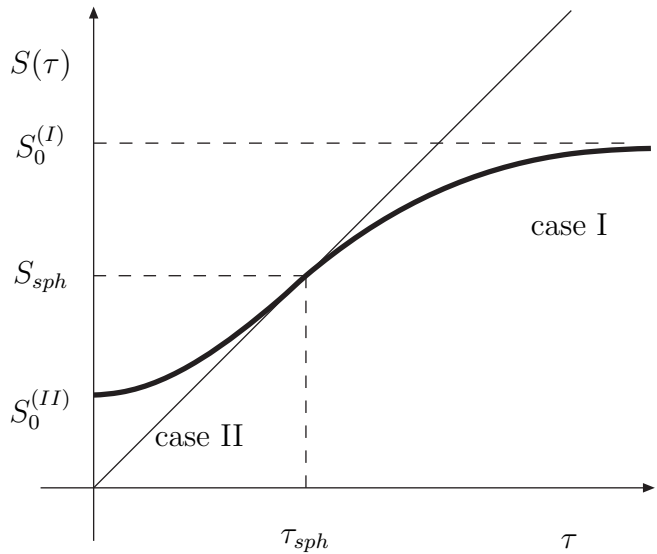
\begin{figure}
\begin{center}
\begin{picture}(230,200)(0,0)
\LongArrow(0,20)(230,20)       
\Text(200,0)[]{$\tau$}
\LongArrow(20,0)(20,200)       
\Text(0,180)[]{$S(\tau)$}
\Line(20,20)(200,200)	       
\DashLine(100,20)(100,100){5}  
\Text(100,0)[]{$\tau_{sph}$}
\DashLine(20,100)(100,100){5}  
\Text(0,100)[]{$S_{sph}$}
\Text(200,120)[]{case I}
\DashLine(20,150)(220,150){5}  
\Text(0,150)[]{$S^{(I)}_0$}
\Text(60,40)[]{case II}
\Text(0,50)[]{$S^{(II)}_0$}
\SetWidth{2}
\Curve{(20,55)(30,56)(90,91)(100,100)(110,108)(200,146)%
(210,147)(230,148)}
\end{picture}
\end{center}
\caption{{\protect\small Two possible types of smooth behaviour of the
periodic instanton action $S$ as a function of its period $\tau$.  The
straight line represents the sphaleron with the action $S=E_{sph}\tau$. }}
\label{S(T)-models}
\end{figure}
These two types of behavior at $E\ll E_{sph}$ can be smoothly
continued into the region $E\sim E_{sph}$ as shown in
Fig.\ref{S(T)-models}. In the case I the period $\tau$ is always
greater than $\tau_{sph}$; the action monotonically increases from
$S_{sph}\equiv E_{sph}\tau_{sph}$ to $S_0$ as the period changes from
$\tau_{sph}$ to $\infty$ (energy changes from $E_{sph}$ to zero). The
curve $S(\tau)$ is convex upward,
\[
{ \partial^2 S \over \partial \tau^2 } 
= \left({ \partial^2 W \over \partial E^2 }\right)^{-1} = \dpar{E}{\tau} 
< 0 . 
\]
Correspondingly, the saddle point (\ref{b=T}) is a {\em maximum} of
the exponent in \eq{T-rate} and saturates the thermal rate at
temperatures $T = 1/\beta < 1/\tau_{sph}$, so that one has 
\begin{eqnarray*}\label{T-rate-via-instanton}
\Gamma(\beta) &\sim & \exp \bigl( -S(\beta) \bigr) 
~~~~~ \mbox{~at $T < 1/\tau_{sph}$}, \\ 
\Gamma(\beta) &\sim & \exp (-\beta E_{sph})
~~~~~ \mbox{at $T > 1/\tau_{sph}$}.  
\end{eqnarray*}	
Note that this case requires $S_0>S_{sph}$. 

In the case II the period is always less than $\tau_{sph}$. The action
monotonically decreases when the period changes from $\tau_{sph}$ to zero
(energy changes from $E_{sph}$ to zero). The curve $S(\tau)$ is convex
downward,
\[
{ \partial^2 S \over \partial \tau^2 } 
= \left({ \partial^2 W \over \partial E^2 }\right)^{-1} = \dpar{E}{\tau} 
> 0 . 
\]
In this case, the saddle point (\ref{b=T}) is a {\em minimum}; the
integral in \eq{T-rate} is saturated either by $E=0$ or by
$E=E_{sph}$, depending on the temperature: 
\begin{eqnarray*}
\Gamma & \sim & \exp(-S_0) ~~~~~ \mbox{at~ $T <
E_{sph}/S_0$}, \\ 
\Gamma & \sim & \exp(-\beta E_{sph}) ~~~~~ \mbox{at~ $T >
E_{sph}/S_0$} .
\end{eqnarray*}	
This case requires $S_0<S_{sph}$. 

Do these two cases comprise all possible types of behaviour?  The
answer is {\em no}. It is easy to find models where neither case I nor
case II can realize. In fact, the Electroweak Theory is one of the
examples. It has no localized instantons and shows low energy
behaviour of the periodic instanton as in the case II
\cite{KRT-periodic}. However, as was first noted in
ref.\cite{HMT-Pioneers}, the numerical analysis of the electroweak
sphaleron \cite{Yaffe} implies that at $M_H > 4M_W$ one has $S_0 >
S_{sph}$, which is not consistent with the case II\footnote{This
anomaly has nothing to do with deformed sphaleron which appears only
at $M_H \simeq 12M_W$}.  This situation is rather common and is due to
the fact that low and high energy limits of periodic instanton are
determined by different parts of the potential and hence are not tied
to each other. Below we present two examples of models where the
periodic instanton can be found explicitly and shows the behaviour
more complicated than in the cases I and II of Fig.\ref{S(T)-models}.

\paragraph{A quantum mechanical model.}
Consider one-dimensional quantum mechanical model with the Euclidean
action
\[
	S_E = {1 \over \lambda} 
		\int d\tau \left( \half \dot\phi^2 + V(\phi) \right). 
\]
The potential $V(\phi)$ is chosen in the form 
\[
	V(\phi) = { 4 + \alpha \over 12 }
		  - \half \phi^2
		  - { \alpha \over 4 } \phi^4
		  + { \alpha+1 \over 6 } \phi^6 .
\]
When $\alpha > -1$ it is of the double well type,
i.e.\ it has two minima at $\phi = \pm 1$
separated by the barrier of the height
\[
	 E_{sph} = { 4 + \alpha \over 12 \lambda }. 
\]
The main difference with the standard (quartic) case is that the walls
of the wells are steeper.

This model possesses exact zero-energy instanton describing quantum
tunneling between two wells, in full analogy with conventional double
well model. Therefore, at $E\to 0$ the period of the periodic
instanton goes to infinity (this is easy to check by direct
calculation, see \eq{QM-solution} below).  On the other hand, in the vicinity of
the sphaleron one finds
\[
	S(\tau) = E_{sph} \tau + { 4\pi\alpha \over 3 \lambda }
			\left( 1 - { \tau \over 2\pi } \right)^2 ,
\]
so that at $\alpha>0$ the periodic instanton has period {\em smaller}
than $\tau_{sph} = 2\pi$ and the value of action {\em larger} than the
corresponding sphaleron value.

In order to see how the two limiting cases match together one has to
find the periodic instanton at intermediate energies. In
quantum-mechanical model this can be done in quadratures,
\begin{eqnarray}\label{QM-solution}
S(\tau) &=& E\tau + {1 \over \lambda} \oint dx \sqrt {2(V(x) - \lambda E)}
							\nonumber\\
\tau    &=& \oint { dx \over \sqrt {2(V(x) - \lambda E)}} 
							\nonumber
\end{eqnarray}
The function $S(\tau)$ for $\alpha=1.5$ is shown
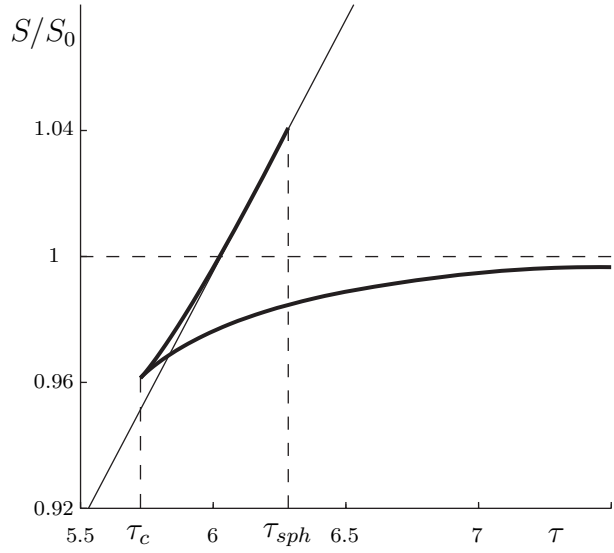
\begin{figure}
\begin{center}
\begin{picture}(210,200)(0,0)
\LinAxis(10,10)(210,10)(4,1,2,0,0.5)
\Text(190,0)[]{$\tau$}
\Text(10,0)[]{$\scriptstyle 5.5$}
\Text(60,0)[]{$\scriptstyle 6$}
\Text(110,0)[]{$\scriptstyle 6.5$}
\Text(160,0)[]{$\scriptstyle 7$}
\LinAxis(10,10)(10,200)(4,1,-2,1,0.5)
\Text(-4,190)[]{$S/S_0$}
\Text(0,10)[]{$\scriptstyle 0.92$}
\Text(0,58)[]{$\scriptstyle 0.96$}
\Text(0,106)[]{$\scriptstyle 1$}
\Text(0,154)[]{$\scriptstyle 1.04$}
\SetOffset(10,10)
%
%
\DashLine(0,94.99)(200,94.99){5}
\DashLine(22.6279,0)(22.6279,49.1891){5}
\DashLine(78.3185,143.523)(78.3185,0){5}
\Line(3,0)(103,190)	       
\Text(22,-10)[]{$\tau_c$}
\Text(78.3185,-10)[]{$\tau_{sph}$}
\SetWidth{1.5}
\Curve{%
(78.3185,143.523)
(74.8876,136.81)
(62.9873,114.147)
(53.3997,96.715)
(45.6564,83.3006)
(39.1585,72.6133)
(34.1418,64.8012)
(30.1357,58.9138)
(27.017,54.5926)
(24.9051,51.8535)
(23.4694,50.1166)
(22.6279,49.1891)
}
\Curve{%
(22.6279,49.1891)
(22.9937,49.6076)
(24.1846,50.7487)
(26.2444,52.5542)
(29.1263,54.8162)
(33.3658,57.7933)
(38.826,61.1432)
(46.3277,65.1133)
(56.088,69.4528)
(69.3053,74.135)
(88.3929,79.2961)
(118.397,84.8318)
(182.843,90.8642)
(200,91)
}
\end{picture}
\end{center}
\caption{{\protect\small The periodic instanton action as
a function of its period in the steep double well model.}}
\label{S(T)-QM}
\end{figure}
in \fig{S(T)-QM}. The periodic instanton indeed exists in some range
$\tau<\tau_{sph}$, but at a critical value of period $\tau_c < \tau_{sph}$ the
curve $S(\tau)$ has a {\em wedge\/} and turns back to large periods. The
branches of the wedge are tangent at the bifurcation point, as follows
from energy continuity and \eq{E=dS/dT}. The curve is convex upward
on the lower branch and downward on the upper one.

\paragraph{$\sigma$-model.}
A field theory model with similar behavour of the periodic instanton
can be constructed from the massive $O(3)$ $\sigma$-model of
ref.\cite{MottWipf} by adding the Skyrme term.  The model is defined
by the Euclidean action
\begin{equation}\label{Skyrme-action}
	S_E = { 1 \over g^2}
	\int d^2 x 	\Bigl\{
			\half n^a_\mu n^a_\mu
			+ (1 + n_3)
			+ { \lambda \over 8 } n^{ab}_{\mu\nu} n^{ab}_{\mu\nu}
			\Bigr\} ,
\end{equation}
where the $n^a$ is a unit vector in 3-dimensional space $a=1,2,3$,
$n^a_\mu = \partial_\mu n^a$ and $n^{ab}_{\mu\nu} = n^a_{[\mu}
n^b_{\nu]}$.  Without the Skyrme term (i.e.\ at $\lambda = 0$), the
massive $\sigma$-model does not have localized instantons and falls
into the class II~\cite{HMT-Pioneers}.  The approximate solutions,
constrained instantons, are unstable and shrink to zero size. At
$\lambda\neq 0$, the size of the constrained instanton stabilizes at
the value of order $(\lambda \log\lambda)^{1/4}$ and the
model~\eqref{Skyrme-action} acquires a localized instanton
solution. It can be found numerically~\cite{Tchrakian}. The double
instanton action at $\lambda \ll 1$ is
\[
	S_0 = { 8\pi \over g^2 } +
	O({\lambda^{1/2} \over g^2}) .
\]
Thus, at $\lambda\neq 0$ one expects the low energy behavoiur as in
the case I. 

The Skyrme term does not affect static configurations, so the
sphaleron solution coincides with that of $\lambda=0$ model, where its
analytical form is \cite{MottWipf}
\[
	n^a = \left(
	 -2 {\sh(x) \over \ch^2(x) },
	0,
	-1 + {2 \over \ch^2(x)}
	      \right) .
\]
The energy of the sphaleron equals
\[
	E_{sph} = { 8 \over g^2 } .
\]

The Skyrme term distorts the matrix of the second derivatives of
the action, so the negative sphaleron eigenmode and, hence,
the critical period depend on $\lambda$. It still can
be found analytically, 
\begin{eqnarray}
u_-^a &=& \delta_2^a { 1 \over \ch^\alpha (x) },\\
\omega_- &=& (\alpha^2-1)^{1/2} , \nonumber
\end{eqnarray}
where $\alpha$ is the positive root of the equation
\[
(1+4\lambda)\alpha^2 + \alpha - 4\lambda - 6=0.
\]
The action of the sphaleron at the critical period is
\[
	S_{sph} = E_{sph}\tau_{sph} = 
	{ 16\pi \over \sqrt{3} g^2 } + O({\lambda \over g^2}) .
\]
Note that $S_0<S_{sph}$, which is not compatible with the case I.

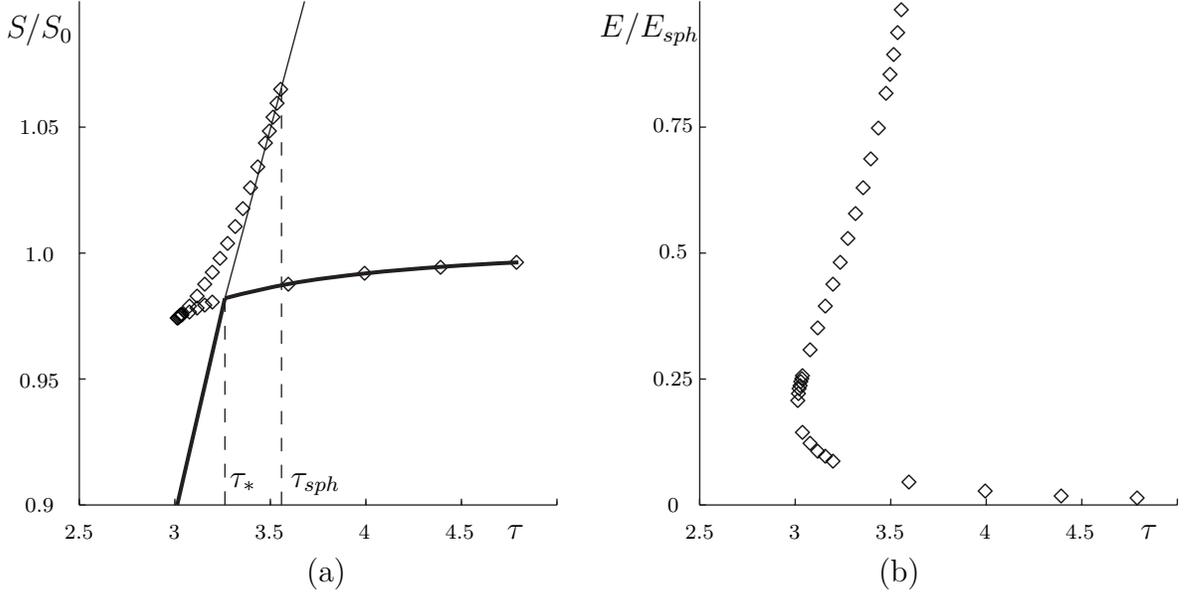
\begin{figure}
\hbox to\textwidth{%
\hbox{%
\begin{picture}(195,200)(0,0)
\LinAxis(15,10)(195,10)(5,1,2,0,0.5)
\Text(180,0)[]{$\tau$}
\Text(15,0)[]{$\scriptstyle 2.5$}
\Text(51,0)[]{$\scriptstyle 3$}
\Text(87,0)[]{$\scriptstyle 3.5$}
\Text(123,0)[]{$\scriptstyle 4$}
\Text(159,0)[]{$\scriptstyle 4.5$}
%
\LinAxis(15,10)(15,200)(4,1,-2,1,0.5)
\Text(0,190)[]{$S/S_0$}
\Text(0,10)[]{$\scriptstyle 0.9$}
\Text(0,57)[]{$\scriptstyle 0.95$}
\Text(0,105)[]{$\scriptstyle 1.0$}
\Text(0,153)[]{$\scriptstyle 1.05$}
\SetOffset(15,10)
\Line(55,78)(85,190)
\DashLine(76.32,157.17)(76.32,0){5}
\Text(80,7)[lb]{$\tau_{sph}$}
\DashLine(55,78)(55,0){5}
\Text(57,7)[lb]{$\tau_*$}
\def\ThePoint{$\diamond$}
\Text(76.32,157.17)[]{\ThePoint}
\Text(74.88,151.703)[]{\ThePoint}
\Text(73.44,146.485)[]{\ThePoint}
\Text(72,141.502)[]{\ThePoint}
\Text(70.56,136.738)[]{\ThePoint}
\Text(67.68,127.819)[]{\ThePoint}
\Text(64.8,119.644)[]{\ThePoint}
\Text(61.92,112.147)[]{\ThePoint}
\Text(59.04,105.275)[]{\ThePoint}
\Text(56.16,98.9815)[]{\ThePoint}
\Text(53.28,93.2329)[]{\ThePoint}
\Text(50.4,88.0009)[]{\ThePoint}
\Text(47.52,83.2658)[]{\ThePoint}
\Text(44.64,79.0177)[]{\ThePoint}
\Text(41.76,75.2615)[]{\ThePoint}
\Text(38.88,72.0363)[]{\ThePoint}
\Text(38.592,71.7473)[]{\ThePoint}
\Text(38.304,71.4654)[]{\ThePoint}
\Text(38.016,71.1911)[]{\ThePoint}
\Text(37.728,70.9253)[]{\ThePoint}
\Text(37.44,70.6691)[]{\ThePoint}
\Text(37.152,70.4253)[]{\ThePoint}
\Text(165.6,91.5303)[]{\ThePoint}
\Text(136.8,89.8682)[]{\ThePoint}
\Text(108,87.3767)[]{\ThePoint}
\Text(79.2,83.4372)[]{\ThePoint}
\Text(50.4,76.4389)[]{\ThePoint}
\Text(47.52,75.4043)[]{\ThePoint}
\Text(44.64,74.2607)[]{\ThePoint}
\Text(41.76,72.976)[]{\ThePoint}
\Text(38.88,71.4862)[]{\ThePoint}
\SetWidth{1.5}
\Curve{%
(165.6,91.5303)
(136.8,89.8682)
(108,87.3767)
(79.2,83.4372)
(55,78)
}
\Line(37,0)(55,78)
\end{picture}
}\hfill\hbox{%
\begin{picture}(190,200)(0,0)
\LinAxis(10,10)(190,10)(5,1,2,0,0.5)
\Text(180,0)[]{$\tau$}
\Text(10,0)[]{$\scriptstyle 2.5$}
\Text(46,0)[]{$\scriptstyle 3$}
\Text(82,0)[]{$\scriptstyle 3.5$}
\Text(118,0)[]{$\scriptstyle 4$}
\Text(154,0)[]{$\scriptstyle 4.5$}
\LinAxis(10,10)(10,200)(4,1,-2,1,0.5)
\Text(-9,190)[]{$E/E_{sph}$}
\Text(0,10)[]{$\scriptstyle 0$}
\Text(0,58)[]{$\scriptstyle 0.25$}
\Text(0,106)[]{$\scriptstyle 0.5$}
\Text(0,154)[]{$\scriptstyle 0.75$}
\SetOffset(10,10)
\def\ThePoint{$\diamond$}
\Text(76.32,187.094)[]{\ThePoint}
\Text(74.88,178.417)[]{\ThePoint}
\Text(73.44,170.341)[]{\ThePoint}
\Text(72,162.782)[]{\ThePoint}
\Text(70.56,155.672)[]{\ThePoint}
\Text(67.68,142.586)[]{\ThePoint}
\Text(64.8,130.744)[]{\ThePoint}
\Text(61.92,119.902)[]{\ThePoint}
\Text(59.04,109.873)[]{\ThePoint}
\Text(56.16,100.503)[]{\ThePoint}
\Text(53.28,91.6612)[]{\ThePoint}
\Text(50.4,83.2176)[]{\ThePoint}
\Text(47.52,75.0304)[]{\ThePoint}
\Text(44.64,66.9084)[]{\ThePoint}
\Text(41.76,58.5116)[]{\ThePoint}
\Text(38.88,48.8655)[]{\ThePoint}
\Text(38.592,47.7103)[]{\ThePoint}
\Text(38.304,46.4793)[]{\ThePoint}
\Text(38.016,45.1428)[]{\ThePoint}
\Text(37.728,43.6474)[]{\ThePoint}
\Text(37.44,41.8745)[]{\ThePoint}
\Text(37.152,39.3981)[]{\ThePoint}
\Text(165.6,2.28029)[]{\ThePoint}
\Text(136.8,3.35559)[]{\ThePoint}
\Text(108,5.12902)[]{\ThePoint}
\Text(79.2,8.41226)[]{\ThePoint}
\Text(50.4,16.4998)[]{\ThePoint}
\Text(47.52,18.1276)[]{\ThePoint}
\Text(44.64,20.1692)[]{\ThePoint}
\Text(41.76,22.9195)[]{\ThePoint}
\Text(38.88,27.3144)[]{\ThePoint}
\end{picture}
}}
\hbox to\textwidth{%
\vrule height18pt depth0pt width0pt
\hbox to 0.5\textwidth{\hfil (a)\hfil}%
\hbox to 0.5\textwidth{\hfil (b)\hfil}}
\caption{{\protect\small (a) The action of the periodic instanton
($\diamond$) in the $\sigma$-model with the Skyrme term, as a function
of its period. Straight line represents the sphaleron. The behaviour
of $S(\beta)=-\ln(\Gamma(\beta))$ is shown by fat line. (b) The energy
of the same periodic instanton as a function of the period.}}
\label{S(T)-Skyrme}
\end{figure}

The periodic instanton can be found numerically. The results of the
calculation at $\lambda=0.001$ are presented in \fig{S(T)-Skyrme}.
As in the case of the steep double well model, there is a wedge at $\tau
< \tau_{sph}$. 

In both examples, the function $S(\tau)$ is multi-valued, and the
period of the periodic instanton is not a monotonic function of
energy. This does not affect qualitatively the probability of
tunneling at fixed energy. In both models $W(E)$ is a monotonically
increasing function of energy; the wedge in the action shows up as
zero of second derivative of $W(E)$. However, the existence of the
wedge has non-trivial consequences for the thermal rate
$\Gamma(\beta)$. The lower branch of the periodic instanton dominates
$\Gamma(\beta)$ at temperatures $T < T_*=1/\tau_*$, where $\tau_*$ is
determined by the intersection of the lower branch with the line
$E_{sph}\tau$. Note that $T_* > T_{sph}$. At higher temperatures, the
rate is controlled by the sphaleron,
\begin{eqnarray*}\label{T-rate-via-instanton-3}
\Gamma(\beta) &\sim & \exp \bigl( -S(\beta) \bigr) 
~~~~~ \mbox{~at $T < 1/\tau_*$}, \\ 
\Gamma(\beta) &\sim & \exp (-\beta E_{sph})
~~~~~ \mbox{at $T > 1/\tau_*$}.  
\end{eqnarray*}	
Thus, in a sense, the two examples considered above are
intermediate between cases I and II.

\paragraph{General properties of periodic instanton bifurcations.}

The wedge in the action that was found in the above examples is likely
to be rather common phenomenon. Moreover, it may be considered as a
particular kind of bifurcations of the periodic instanton solution
under the change of parameters (period $\tau$ here) and should be
treated on equal footing with the bifurcation point at
$\tau=\tau_{sph}$ where the periodic instanton merges to the
sphaleron. Since the behaviour of solutions near bifurcation points is
governed by global properties of the action, its general features can
be analyzed without referring to particular model.

Consider the hessian of the action for a given periodic instanton
$I$. In general, the periodic instanton is not a local minimum of the
action, so the hessian has a finite number $\mu(I)$ of negative
eigenvalues. Generally, it has also zero modes, corresponding to
global symmetries of the model, but we consider them pre-eliminated,
so that for generic values of parameters the hessian is supposed to be
non-degenerate and, hence, $\mu(I)$ is the Morse index of the solution
$I$.  The determinant of the hessian gives pre-exponential factor in
\eq{Semiclassic-P}, which is ignored throughout the paper. However, it
contains an important piece of information about the leading
approximation. Namely, each negative mode of the hessian contributes
into \eq{Semiclassic-P} a factor $i$.  The path integral
representation for $P(E)$ contains also integrations over discrete
variables which finally reduce to integration over the period $\tau$
\cite{KRT-periodic} (so that \eq{E=dS/dT} is in fact the saddle point
equation). This integration may give another factor $i$. Thus,
the requirement that $P(E)$ is real translates into equation
\begin{equation}\label{Counting-rule-1}
\left[{\rm sign} \dpar{E}{\tau} \right] (-1)^{\mu(I)} = 1 ,
\end{equation}
were we made use of the relation $\partial^2 S/\partial^2\tau =
\partial E/\partial \tau$. Eq.(\ref{Counting-rule-1}) should be
treated as a selection rule for solutions which can contribute into
$P(E)$. According to this rule, the periodic instantons of types I and
II should have an odd and even number of negative modes,
respectively. As is shown below, this is indeed the case. Similar
statement applies to the lower and upper branches of the wedge.

When several solutions merge together one can define the degree $d$ of
the bifurcation point as sum of $(-1)^{\mu(I)}$ over all merging
solutions:
\[
	d = \sum_I (-1)^{\mu(I)} .
\]
This number is a topological invariant, which has the same value
before and after bifurcation.  Actually, the low dimensional geometric
intuition implies that the number of negative modes on merging
branches monotonically increases with their action.

As an example consider the wedge discussed above. Before the
bifurcation point ($\tau<\tau_c$) there are no solutions, so $d=0$.
Therefore, at $\tau>\tau_c$ the two emerging solutions must obey
$\mu(I_1) + \mu(I_2) = \mbox{odd}$ (and, therefore, should have
opposite signs of $\partial E/\partial \tau$), i.e.\ the lower and
upper branches should have an odd and even number of negative modes,
respectively. This fact can be easily checked in quantum mechanics by
direct numerical diagonalization of the discretized hessian; for field
models this statement is less trivial.

Another application of the conservation of $d$ is the bifurcation of
the sphaleron at $\tau=\tau_{sph}$. The sphaleron solution has exactly
one static negative mode. Consider time dependent modes of the hessian
\[
	H = - \partial^2_\tau + H_{stat}
\]
in the vicinity of the sphaleron.  Here $H_{stat}$ is the
time-independent part of the hessian. The static eigenmodes of the
sphaleron diagonalize $H_{stat}$, 
\[
	H_{stat} = \sum \omega^2_n u_n \otimes u_n. 
\]
Therefore, the full hessian has eigenmodes 
\[
	U_{mn}(t,x) = \cos { 2\pi m t \over \tau } u_n(x)
\]
and eigenvalues 
\[
\Lambda_{mn} = \omega^2_n + \left({ 2\pi m \over \tau }\right)^2 .
\]
The static negative mode of the sphaleron corresponds to $n=0$, 
so that $\omega^2_0 = - \omega_-^2 < 0$. We see that at 
$\tau < \tau_{sph} = 2\pi/\omega_-$ the full hessian has
exactly one negative mode $\Lambda_{00} = - \omega_-^2$, but at
$\tau > \tau_{sph}$ another negative mode appears with the 
eigenvalue 
\[
\Lambda_{10} = - \omega^2_- + \left({ 2\pi \over \tau }\right)^2 <0. 
\]
In order to satisfy the conservation of $d$ at $\tau=\tau_{sph}$ we
have to assume the existence of at least {\em two} periodic instanton
solutions, both merging or splitting when $\tau$ passes through
$\tau_{sph}$.  Moreover, in the case of splitting these solutions
should have one negative mode each, while in the case of merging two
negative modes each. We indeed have exactly two periodic instantons
differing by translation in time by $\tau/2$, so this analysis not
only proves the existence of periodic instanton in the vicinity of the
sphaleron (this fact is non-trivial in models with multiple degrees of
freedom), but also predicts the number of their negative modes.

To summarize this discussion, the behaviour of periodic instantons is
governed by a set of simple rules: i) The action always increases with
the period, as follows from \eq{E=dS/dT}. ii) The energy monotonically
decreases as moving along the curve of periodic instantons from the
sphaleron. iii) The curve is convex upward or downward at $\partial
E/\partial \tau$ negative or positive, respectively. iv) The number of
negative modes is correlated with the sign of $\partial E/\partial
\tau$ so as to obey \eq{Counting-rule-1}. v) At bifurcation points
the degree $d$ must conserve and the solution with larger action
should have more negative modes. 

\begin{figure}
\hbox to\textwidth{%
\hbox{%
\begin{picture}(200,170)(0,0)
\LongArrow(0,20)(200,20)       
\Text(190,0)[]{$\tau$}
\LongArrow(20,0)(20,170)       
\Text(0,160)[]{$S(\tau)$}
\Line(20,20)(170,170)	       
\DashLine(70,20)(70,70){5}     
\Text(70,0)[]{$\tau_{sph}$}
\DashLine(160,20)(160,130){5}  
\Text(160,0)[]{$\tau_{c}$}
\DashLine(20,70)(70,70){5}  
\Text(0,70)[]{$S_{sph}$}
\Text(0,100)[]{$S_0$}
\SetWidth{2}
\Curve{(20,100)(40,101)(155,128)(160,130)}
\Text(143,112)[]{$1$}
\Text(70,115)[]{$2$}
\Curve{(70,70)(88,86.5)(155,128)(160,130)}
\end{picture}
}\hfill\hbox{%
\begin{picture}(200,170)(0,0)
\LongArrow(0,20)(200,20)       
\Text(190,0)[]{$\tau$}
\LongArrow(20,0)(20,170)       
\Text(0,160)[]{$S(\tau)$}
\Line(70,70)(170,170)	       
\DashLine(70,20)(70,70){5}     
\Text(70,0)[]{$\tau_{sph}$}
\DashLine(106,20)(106,100){5}  
\Text(106,0)[]{$\tau_{*}$}
\DashLine(20,70)(70,70){5}  
\Text(0,70)[]{$S_{sph}$}
\DashLine(20,100)(106,100){5}  
\Text(0,100)[]{$S_0$}
\Curve{(20,100)(40,101)(155,128)(160,130)}
\Curve{(70,70)(88,86.5)(155,128)(160,130)}
\SetWidth{2}
\Line(20,20)(70,70)	       
\Line(106,100)(200,100)  
\Curve{(70,70)(88,86.5)(106,100)}
\end{picture}
}}
\hbox to\textwidth{%
\vrule height18pt depth0pt width0pt
\hbox to 0.5\textwidth{\hfil (a)\hfil}%
\hbox to 0.5\textwidth{\hfil (b)\hfil}}

\caption{{\protect\small (a) Expected behaviour of the periodic
instanton action, $S(\tau)$, in the Electroweak theory at
$4M_W<M_H<12M_W$. Straight line represents the sphaleron. (b)
Corresponding function $S(\beta)$ (fat line) which determines the
thermal rate $\Gamma\sim\exp[-S(\beta)]$.}}
\label{S(T)-EW}
\end{figure}
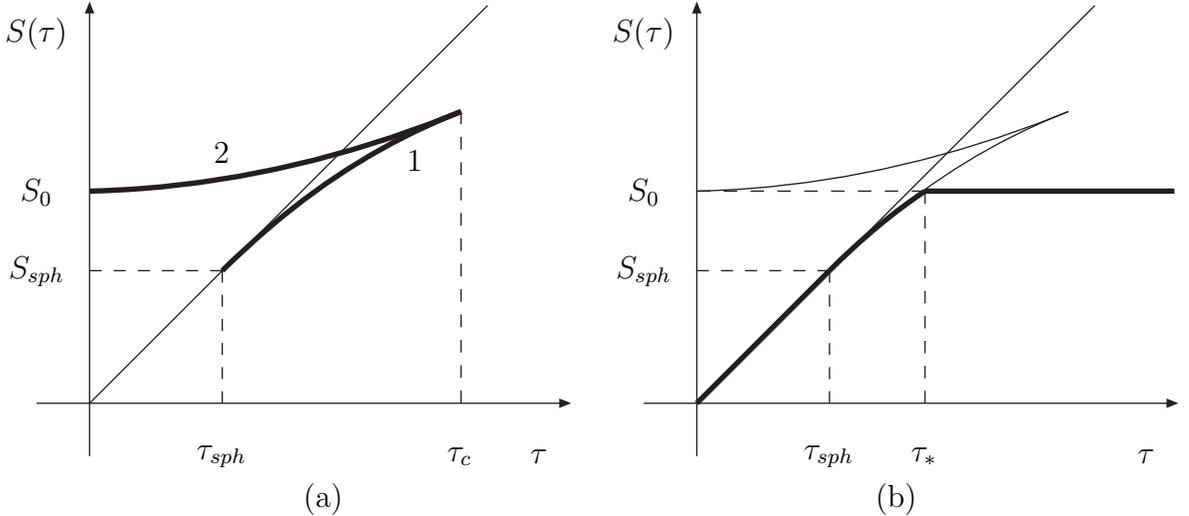
In conclusion, let us apply these rules to the periodic instanton in
the Electroweak Theory at $12M_W>M_H>4M_W$. In this case one has
$S_0>S_{sph}$, while the low energy behaviour of the periodic
instanton follows the case II. The simplest way to reconcile these two
facts is shown in \fig{S(T)-EW}a (if the behaviour of the periodic
instanton near the sphaleron is as in the case II, it must have at
least two wedges). The lower branch (consisting of two identical
periodic instantons which are shifted by $\tau/2$) has one negative
mode, while the upper branch has two negative modes. This has an
implication in the calculation of the thermal rate: at temperatures
$T$ satisfying $\tau_{sph} < 1/T < \tau_*$ the latter is saturated by
the lower branch of periodic instanton, see \fig{S(T)-EW}b. This
picture survives up to $M_H \sim 12M_W$, where the sphaleron itself
bifurcates~\cite{Yaffe}.

\paragraph{Acknowledgements.}

The authors are grateful to V.~A.~Rubakov
for numerous discussions at different stages of this work.
The work is supported in part by Award
No. RP1-187 of the U.S. Civilian Research \& Development Foundation for
the Independent States of the Former Soviet Union (CRDF), and by
Russian Foundation for Basic Research, grant 96-02-17804a.

\end{document}